\documentstyle[12pt,paspconf,psfig]{article}

\def\R{\rm I\kern-.18em R}

\begin{document}

\title{Averaging Inhomogeneous Cosmologies} 
\vskip  -0.8 true cm
\title{-- a Dialogue\altaffilmark{1}}

\author{Thomas Buchert}
\affil{Theoretische Physik, Ludwig--Maximilians--Universit\"at, 
Theresienstr. 37, D--80333 M\"unchen, F.R.G.,\\
e--mail: buchert@stat.physik.uni--muenchen.de}

\altaffiltext{1}{Proc. 2$^{\rm nd}$ SFB Workshop on Astro--particle physics, 
Ringberg 1996, Proceedings Series SFB 375/P002 (1997), R. Bender, T. Buchert,
P. Schneider and F.v. Feilitzsch (eds.).}

\begin{abstract}

\noindent
The averaging problem for inhomogeneous cosmologies is discussed in the form
of a disputation between two cosmologists, one of them (RED) advocating the 
standard model, the other (GREEN) advancing some arguments against it.  
Technical explanations of these arguments as well as the conclusions 
of this debate are given by BLUE.   
\end{abstract}



\section{The conjecture about the average flow}

The standard model which, on some large scale, is defined as a homogeneous and isotropic solution
of Einstein's equations for gravitationally interacting matter,
has proved to be remarkably robust against various observational challenges
especially of the recent past. It is this robustness together with a list 
of theoretical and observational arguments which makes it hard to see any
need for an alternative\altaffilmark{2} to the standard model. Nevertheless, there exist
some simple arguments which let the standard model appear dogmatic and 
a replacement overdue, while most scientific activity in the field is 
directed towards a consolidation of the standard model. It is fair to say that 
most of the work, which is directed towards consolidation,
is already based implicitly on the assumption that the standard model gives 
the correct picture.

\altaffiltext{2}{`alternative' is {\it not} meant in the sense of invoking 
physical laws other than general relativity and making generalizations  
which depart from the standard kinematical properties of an, on average, 
homogeneous--isotropic universe. Rather we think at 
improving on the 
standard model in its presently employed form. Compare also the discussion
in (Ellis et al. 1997).}

\smallskip
I here scetch a possible dialogue which we can watch without risk of
being biassed by some prejudice: we have two people who try to defend their
points of view and both of them might be biassed, but both advance arguments 
which can be proved or disproved. This dialogue is mirrored in an ongoing 
debate in the field of astrostatistics about the existence of 
evidence for a scale
of homogeneity (e.g., Davis 1996 and Pietronero 1996), 
a subject which is also dealt with 
in several contributions to this volume 
(see: Kerscher et al., Mart\'\i nez, Sylos Labini et al.) and was the
subject of a panel discussion held during the meeting. 

\smallskip
Let us start with the advocate of the standard model RED:
``The (large--scale) standard model\altaffilmark{3} is a solution of Friedmann's equations:
$$
3\left( {{\dot a}\over a}\right)^2 - 8\pi G
{M\over a^3} + {k\over 2 a^2} 
- \Lambda\; =\; 0\;\;;\;\;\varrho_H a^3 = :M = const.
\;\;\;,\eqno(1a,b)
$$
where $\Lambda$ is the cosmological constant, $k$ is related to the
constant curvature of the model at time $t_0$, $\varrho_H (t)$ the value
of the homogeneous density at time $t$, and $a(t)$ is the scale function of
the isotropically expanding (or contracting) universe.''
Also: ``This model is unstable to perturbations in the density field
and/or the velocity field, respectively'', which
is the well--known content of gravitational instability; we may
call this property {\it local gravitational instability}. 
In spite of this instability, 
RED supports the following conjecture about the global properties of the 
Universe (a statement which will come again 
in various refined versions later on):

\altaffiltext{3}{For convenience we restrict the discussion to the matter
dominated era.}
 
\bigskip

\noindent
{\bf Conjecture} (Version 1): ``The Universe can be {\it approximated} 
by the standard model,
if {\it averaged} on some large scale $L$, e.g. for the inhomogeneous
density field we would have: $\langle \varrho (\vec x,t)\rangle_L 
(t) = \varrho_H (t)$ for all times.''

\bigskip\noindent
Here, we may think for simplicity that the brackets 
$\langle {\cal A} \rangle$ denote Euclidean 
spatial averages of some tensor field as a function of Eulerian coordinates and
time, ${\cal A} (\vec x,t)$. The Newtonian case serves as a good
illustration. Below, we shall
explain that the arguments carry over to Riemannian spaces and general 
relativity.

\smallskip\noindent
GREEN replies: ``I don't expect that spatial averaging and time evolution
commute as a result of the nonlinearity of the basic system of equations.''

\noindent
BLUE explains:
``If we average the cosmological fields of density and velocity at some 
initial time $t_0$ (at recombination\altaffilmark{4})
\altaffiltext{4}{This does, however, not imply 
that the `backreaction' discussed hereafter will be unimportant at earlier times; it
may even be relevant in the early universe as pointed out by Mukhanov et al.
(1997).} and use these average values 
(which are remarkably isotropic according to the microwave background 
measurements) as initial 
data of a homogeneous--isotropic solution, then, e.g., the value of $\varrho_H (t)$ 
at time $t$ is expected to differ from the average field $\langle \varrho  
\rangle_L (t)$ of the inhomogeneous initial data evolved to the time $t$.
This has been particularly emphasized by Ellis (1984).'' 
Indeed, GREEN is right: the explanation of non--commutativity is given by
BLUE in terms of the {\it commutation rule} for the expansion scalar defined as
the divergence of the velocity field $\theta=\nabla \cdot \vec v$
(Buchert, Ehlers 1997):
$$
{d\over dt} \langle\theta \rangle_{D} - \langle{d\over dt}\theta \rangle_{D} \;=\; 
\langle\theta^2 \rangle_{D} - \langle\theta\rangle_{D}^2 \;\;\;.\eqno(2)
$$
``Equation (2) shows that, on any spatial domain $\cal D$, the evolution of the
average quantity differs from the averaged evolved one, the difference being
given by a fluctuation term.
For details and discussions of what follows see (Buchert 1996 and 
Buchert \& Ehlers 1997);
for the relation to dynamical models see (Ehlers \& Buchert 1997).''

Also:
\smallskip
``Equation (2) only assumes mass conservation, i.e., we follow a tube of 
trajectories so that the
mass in the spatial averaging domain $\cal D$ is conserved in time. This is
a sensible assumption, since we want to extend the spatial domain to the whole
universe later on.''

\section{A generalized expansion law}

\noindent
BLUE goes further by specifying the local dynamical law for the expansion 
scalar. This is furnished by Raychaudhuri's equation: ``Introducing a 
scale factor via the volume $V=|{\cal D}|$, $a_{\cal D} (t):=V^{1/3}$ (so, we do not care 
about the shape of the spatial domain $\cal D$; it may expand anisotropically),
Raychaudhuri's equation for $\dot\theta$ may be inserted into the 
commutation rule (2) resulting in the {\it generalized Friedmann equations}:
$$
3{\ddot a_{\cal D} \over a_{\cal D}} + 4\pi G \varrho_{\rm eff} -
\Lambda = 0 \;\;,\;\;
{d\over dt}\langle\varrho\rangle_{\cal D} + 
3 {\dot a_{\cal D} \over a_{\cal D}}\langle\varrho\rangle_{\cal D} = 0\;\;,\eqno(3a,b)
$$
with the {\it effective} source term which involves averages over fluctuation
terms of the expansion, the shear scalar $\sigma$ and the vorticity scalar
$\omega$: 
$$
4\pi G\varrho_{\rm eff} : = 4\pi G\langle\varrho\rangle_{\cal D} - 
\langle{\cal Q}\rangle_{\cal D}\;\;,\eqno(3c)
$$
$$
\langle{\cal Q}\rangle_{\cal D} = {2\over 3}
\langle\left(\theta - \langle\theta\rangle_{\cal D}\right)^2\rangle_{\cal D} 
- 2\langle\sigma^2 -\omega^2 \rangle_{\cal D}\;\;.\eqno(3d)
$$
Thus, as soon as inhomogeneities are present, they are sources of the 
equation governing the average expansion. They may be negative or positive
giving rise to an additional effective (dynamical) density which we may measure by the
dimensionless ratio 
$$
{\cal B}_D := {\langle{\cal Q}\rangle_{\cal D}\over 4\pi G \langle \varrho 
\rangle_{\cal D}} \;\;\;.
\eqno(4)
$$
For ${\cal B}_D =1$, the source due to 
`backreaction' is, on the averaging domain, equal to that of the averaged matter 
density. The effective density does, in general, not obey a continuity 
equation like the matter density; an effective mass is either produced or
destroyed in the course of structure formation.''

\noindent
RED: ``In principle you are right, but I doubt that the effect is 
quantitatively significant.''

\noindent
GREEN: ``Irrespective of the global relevance of this term, it will play 
an interesting role on scales where its value is non--negligible; for 
dominating shear fluctuations the `backreaction' could fake a `dark matter'
component, since the mass in the standard Friedmann equation will be 
overestimated in this case.''

\section{Inhomogeneous Newtonian cosmologies}
\vspace{-0.3 cm}
\section*{$\;\;\;\;\;\;\;$which are Friedmannian on average}

\noindent
RED is going to advance a strong argument in favour of the standard model:
``In Newtonian theory the `backreaction' term (4) vanishes by averaging over the whole Universe, if the
latter is topologically closed, i.e., compact and without boundary.''

\noindent
Indeed, he succeeds in writing the local term $\cal Q$ as a divergence of some
vector field $\vec\Psi$, ${\cal Q}=\nabla\cdot{\vec\Psi}$, if he assumes 
space to be Euclidean. ``Hence, using Gau{\ss}'s
theorem, we may transform the volume integral in the average $\langle {\cal Q}
\rangle_{\cal D}$ into a surface integral over the boundary $\partial {\cal D}$
of the averaging domain. For compact universes without boundary (e.g., a 
3--torus ${\cal T}$) this surface
integral is zero; we obtain $\langle {\cal Q}\rangle_{\cal T} = 0$ on 
the torus.''

\noindent  
BLUE: ``As a side--result RED proved that the currently employed models for 
large--scale structure, analytical or N--body simulations, are constructed 
correctly: the assumption {\it of periodic boundary conditions} is equivalent
to using as the 3--space a hypertorus, not $\R^3$. 
These models so far have been 
assumed to be Friedmannian on average
{\it by construction} rather than derivation.''

\section{Generalized expansion law in general relativity}

\noindent
GREEN adds a disclaimer: ``The above argument depends on the flatness
of space.''  

\noindent
RED: ``But, as was shown in (Buchert \& Ehlers 1997) the generalized 
Friedmann equations (3)
also hold for irrotational flows in general relativity.''

\noindent
GREEN: ``Yes, but not the fact that the local term ${\cal Q}$ 
can be written as a divergence of some vector field, which makes, besides
spatial curvature, a crucial difference.''

\noindent
BLUE illustrates this last statement by GREEN by some technical explanations:
``If one introduces normal (Gaussian) coordinates $X^i$
and thus foliates spacetime into flow--orthogonal hypersurfaces 
of constant time (this is only possible for irrotational flows),
then Eqs. (3) also hold.
Since Eqs. (3) are equations for scalar quantities, they are manifestly covariant
and will hold in any coordinate system.
The crucial difference to the Newtonian treatment, however, 
is the non--integrability
of inhomogeneous deformations (as defined below): 
Write the metric of the spatial hypersurfaces as
a quadratic form
$g_{ij}=\eta^a_{\;i} \eta^a_{\;j}$ involving the one--forms 
${\bf\eta}^a = \eta^a_{\;i}dX^i$, then it is necessary and sufficient for the
metric to be flat that the one--forms are exact, i.e., ${\bf\eta}^a = 
df^a$, and the coefficients 
$\eta^a_{\;i}$ reduce to a deformation gradient 
with respect to Lagrangian coordinates, 
$df^a_{\;i}\equiv \partial f^a / \partial X_i$; in other words, the coefficient
matrix $\eta^a_{\;j}$ which measures the deformations is integrable. 
The non--integrability of $\eta^a_{\;j}$ implies the non--existence of a vector field 
$\vec\Psi$. Therefore, we cannot shift a volume averaged quantity to
a contribution on the boundary of the averaging domain and the conclusion
on the vanishing of `backreaction' for models with non--Euclidean
space sections cannot be drawn using RED's argument.'' 

\smallskip\noindent
GREEN: ``This remark by BLUE is bad news in so far as we expect the valid theory to be
general relativity on the large scales under consideration.''

\smallskip\noindent
BLUE: ``Again, we can formulate a side--result concerning current models of 
large--scale
structure: If we model structure in, e.g., an undercritical--density  
universe (the total density parameter being $\Omega < 1$), then the model {\it has
to be} interpreted as a Newtonian model. It makes sense to speak about
hypersurfaces of constant negative curvature for the average model, but in 
that case there currently exists no proof that the average model is 
Friedmannian for closed, curved spaceforms. Moreover, it is then not 
even possible to introduce a simple hypertorus topology, since this is 
incompatible with a hyperbolic geometry (compare, e.g., Lachi\`eze--Rey \& 
Luminet 1995).''

\section{Enforcing closure for spatial hypersurfaces}

RED summarizes the preceeding findings and reformulates his statement:
\smallskip

\noindent
{\bf Conjecture} (Version 2): ``On some large scale $L$ we still may approximate
the metric of spatial hypersurfaces by the flat Euclidean metric.
Then, we have two options (which both are connected with the
requirement of periodic boundary conditions):
\smallskip

Either,
 
$\bullet$ We live in a ``small universe'' (Ellis 1971, Ellis \& Schreiber 1986),
i.e., space is genuinely compact without boundary and has a finite size $L^3$. 
This we may call {\it topological closure condition} (Option A).
\smallskip

Or,

$\bullet$ The value of $\langle {\cal Q}\rangle_L$ is numerically negligible 
on the scale $L$. This would support the generally held view and we may call this 
{\it technical closure condition} (Option B), since then Option A is a justified
approximation.''

\smallskip\noindent
GREEN: ``I agree that a compact universe is appealing, since
only then we have some
hope to explore a substantial fraction of its volume; however, it may then
not have flat space sections.''
\smallskip

\section{Is there a compensation of fluctuation terms ?} 

GREEN accepts the flatness of space as a working hypothesis in order not to
complicate the discussion: ``If we don't have Option A (in which case
RED's argument is exact), we have to examine the `backreaction' term 
$\langle{\cal Q}\rangle_{\cal D}$ in more detail quantitatively''.

\noindent
GREEN develops his argument: 
``The terms in $\langle{\cal Q}\rangle_{\cal D}$ have 
positive contributions (vorticity and expansion fluctuations), 
$\langle{\cal Q}^+\rangle_{\cal D}$,
and negative ones (shear fluctuations), 
$\langle{\cal Q}^- \rangle_{\cal D}$'', and ``Each individual fluctuation term
has fixed sign and, thus, does not vanish on {\it any} scale''.

\noindent
BLUE: ``An immediate consequence of the positivity of these terms is that their values 
may decay with scale until we reach a representative volume of the Universe,
but as soon as we have reached this, these terms approach a finite positive
value, even on a scale on which we may assume periodic boundary conditions.''

\noindent
GREEN concludes that ``The requirement of vanishing or smallness of the sum
$\langle{\cal Q}^+\rangle_{\cal D} +\langle{\cal Q}^-\rangle_{\cal D}$
implies a {\it conspiracy} between vorticity, shear and expansion
fluctuations, which is not to be expected a priori.''

\noindent
The final refinement of RED's statement therefore assumes the form:

\bigskip\noindent
{\bf Conjecture} (Version 3): ``On some large scale $L$ we still may approximate
the metric of spatial hypersurfaces by the flat Euclidean metric. 
In general we may not expect that the Universe is genuinely periodic on the
scale $L$. However, on that scale, 
the term $\langle{\cal Q}\rangle_L$ has a negligible value: 

Either, because:

$\bullet$ Each of the terms $\langle{\cal Q}^+\rangle_L$ and 
$\langle{\cal Q}^-\rangle_L$ is numerically small, so that the conspiracy 
assumption does not matter.

Or, because (if the terms are not numerically small):

$\bullet$ The inhomogeneities evolve such that 
$\langle{\cal Q}^+\rangle_{\cal D} +\langle{\cal Q}^-\rangle_{\cal D} 
\rightarrow 0$ for scales approaching $L$ 
{\it and for all times}.''

\bigskip\noindent
BLUE: ``Both options imply assumptions on the initial fluctuation spectrum, and 
both formulate properties of gravitational {\it dynamics} which is, in
principle, testable.''

\noindent
RED: ``I agree that we want, under all circumstances, avoid 
``fine--tuning''; the standard model should be generic for a wide range
of dynamical models for the evolution of inhomogeneities.''

\section{A physical model}

GREEN now argues on the grounds of gravitational dynamics:
``Up until now there is no dynamical model which includes the full 
`backreaction' (apart from perturbative studies which may capture some
of the effect -- see Futamase 1989, 1996, Bildhauer 1990 and 
Bildhauer \& Futamase 1991a,b, as well as Russ et al. 1997);
the main problem to construct such a model is the following: not only
the inhomogeneities affect the global expansion, but also any model for 
the evolution of inhomogeneities will depend on how the average evolves 
in time.'' 

\noindent
BLUE details: 
``The latter is principally known from the linear theory of gravitational 
instability: if the universe expands faster, then the inhomogeneities have a 
harder time to form. Here, we are faced with a nonlinear self--interaction
problem:

\noindent
We may start with a flat Friedmann model as background 
(the average of the Ricci scalar is zero), and some model for the 
inhomogeneities relative to this background. From the inhomogeneous model
we calculate the `backreaction' (this was recently attempted by Russ et al. 1997).
However, if there is a nonvanishing `backreaction' on the global scale $L$,
then this procedure gives us just the first step in the sense of an iteration;
in the second step we would have a curved background (the average of the Ricci
scalar is nonzero), and we would have to construct an inhomogeneous model for a 
curved background including the `backreaction' from the first iteration. 
In turn the second iteration would yield the `backreaction' for this model,
and the full `backreaction' could be calculated after $N$ steps of this 
procedure provided there is convergence to a solution.''

\noindent
GREEN: ``It is clear that we are far from being able to investigate such a model.
For example, in a curved background we can neither use simple periodic boundary 
conditions, nor can we work with the standard Fourier transformation; 
we would have to work with eigenfunctions on curved spaces and would have to
respect the compatibility with some, in general, 
nontrivial topologies.''
\smallskip\noindent
One ``way out'' is to cheat: GREEN bases his further argumentation
on the standard Newtonian model: ``We may use the standard model which is 
mathematically well--defined as the average over a general inhomogeneous
but periodic Newtonian model, and let the box of the simulation extend to 
very large scales. Then, the `backreaction' can be calculated for subensembles
of the simulation box on scales which we consider representative for the 
Universe.'' This possible study will at least give us some quantitative
clues of the effect; it is the subject of an ongoing work 
(Buchert et al. 1997) which BLUE is going to scetch in the next section.

\section{A dynamical approach to cosmic variance}

``Let us assume, in agreement with Conjecture 3 by RED, that the space sections
are Euclidean and that the inhomogeneities can be subjected to periodic boundary
conditions on some large scale. Usually, this scale is set to be around 
$300$Mpc/h, mainly because of limits on CPU power using N--body simulations.
However, a recent analysis of the IRAS 1.2 Jy catalogue (see Kerscher et al., 
this volume)
has demonstrated that fluctuations in the matter distribution do not vanish
on that scale. A mock catalogue of the IRAS sample produced by a simulation with
a box size of $250$Mpc/h enforces the fluctuations to vanish on the periodicity
scale, and the corresponding analysis of the mock sample shows disagreement 
in all moments (except the first) of the matter distribution with the observed data 
(see Kerscher et al., this volume and Kerscher et al. 1997).
This example shows that not only fluctuations in the average density are 
an indication for inhomogeneities, but averages over higher--order moments 
of the density field (e.g. reflected by averaged 
shear fluctuations\altaffilmark{5}) create
huge (phase--correlated), possibly low--amplitude structures.

\altaffiltext{5}{Shear fluctuations are accessible
through observations: the Mark III catalogue of peculiar--velocities offers
this access, which is an especially interesting study, since reconstruction
of the density field is performed just on the scales where a large effect 
is to be expected.}

Thus, even if we do not argue globally about the 
`backreaction problem' (the `backreaction' is zero by construction due to the 
assumed Newtonian description and the periodicity), this effect has to be
seriously considered on scales of current all--sky surveys.
This study entails, from a dynamical point of view, a quantification of 
{\it cosmic variance} within the standard model.

We therefore have to run simulations of a considerably larger spatial extent;
we may use for simplicity 
``truncated Lagrangian schemes''. These schemes have been shown to agree with
N--body results down to scales around the correlation length 
(Melott et al. 1994, Wei{\ss} et al. 1996) and, thus, are
suitable tools to realize boxes of Gigaparsec extent.
For this purpose it may be considered sufficient to use the ``truncated first--order
scheme'' (known as TZA; ``Truncated Zel'dovich Approximation'', Coles et al. 1993).

In a work in progress Buchert et al. (1997) consider two COBE normalized 
cosmogonies, Standard--CDM and a CDM model with cosmological constant.
Both cosmogonies are realized for a box of 1.8 Gpc/h with an effective 
resolution of $3000^3$ Lagrangian fluid elements. 
The simulation box is then subdivided into smaller boxes and the ensemble
average is taken over values of the dimensionless relative `backreaction' (4).
Other quantities like the expected Hubble constant or the expected 
density parameter including `backreaction' are also studied both numerically and
analytically.

\begin{figure}[ht]
\hspace{-0.5cm}\vspace*{-0.8cm}
\psfig{figure=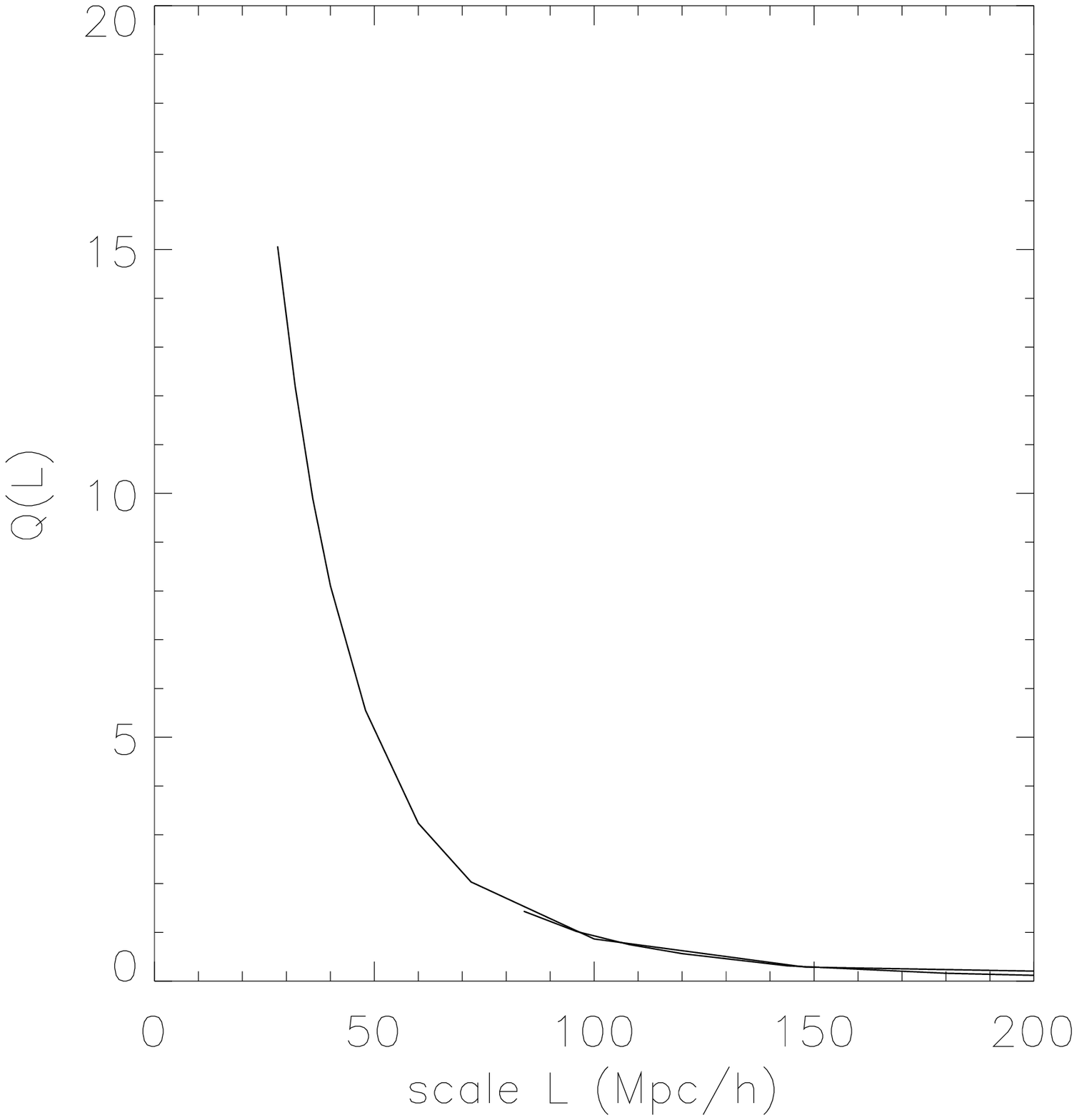,height=6.8cm,width=6.8cm}
\vskip -6.2 true cm
\rightline{\psfig{figure=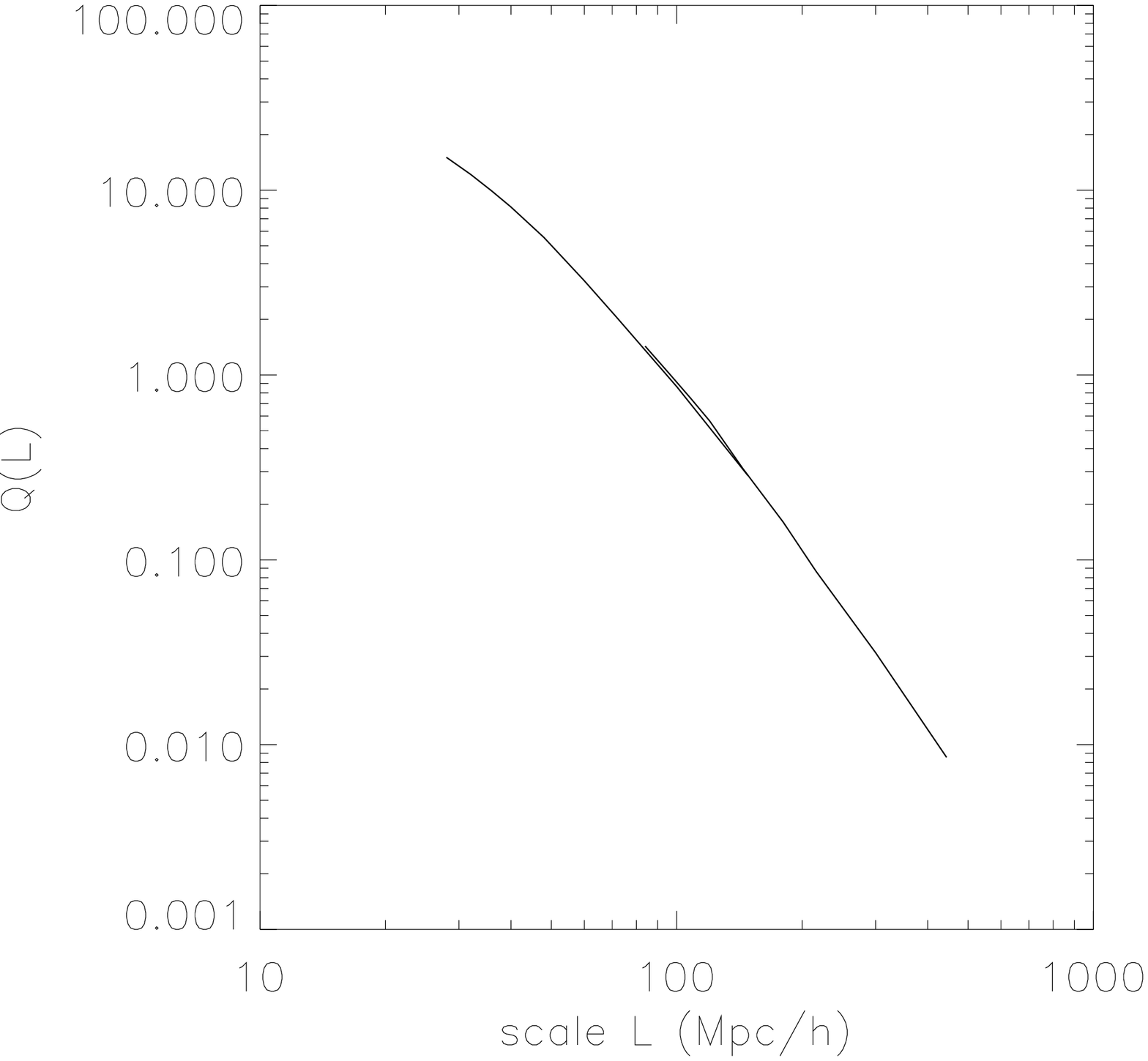,height=7.0cm,width=7.0cm}}


\caption{\small The expected `backreaction', i.e., the 
quantity $|{\cal Q}_{\cal D}|$ averaged over all subensembles on the 
respective scale and over
$8$ random phase realizations, normalized by the background density 
$4\pi G\varrho_H$. Two box sizes ($600$Mpc/h and
$1.8$Gpc/h) of a Standard--CDM model are shown together demonstrating 
that the results of the two runs match. The `backreaction' is represented as
a function of the scale $L \equiv a_{\cal D} = |{\cal D}|^{1/3}$
(measured in Mpc/h) in linear/linear (left panel) and log/log (right panel) 
format.
This dimensionless quantity is still of the 
same order as the actual matter density on scales around $100$Mpc/h.
Normalized in addition by the r.m.s.
density fluctuations, calculated with top--hat smoothing, we get a value
which is almost constant with scale of about 70.} 
\label{fig-1}
\end{figure}

A plot of the absolute value of ${\cal Q}_{\cal D}$, normalized by the global 
mean density, for 
the initial conditions SCDM against scale is shown in Figure 1, which already gives
a clear representation of the scale dependence of the effect under study.
The absolute values are shown here, because 
${\cal Q}_{\cal D}$ might be negative or 
positive in subsamples, and the overall sum is, by assumption, zero.
The absolute value is then an estimate of the expected
`backreaction' on some scale. However, in some subsamples the effect may be
smaller or larger. 

Three preliminary conclusions may be drawn from the first results of this 
study: 

\smallskip\noindent
$\bullet$ The magnitude of the `backreaction' source term is of the same order 
as the mean density and higher
on scales $\lambda < 100$Mpc/h for SCDM. It quickly drops to a $10\%$ effect
on scales of $\lambda \approx 200$Mpc/h.

\smallskip\noindent
$\bullet$ The magnitude of the `backreaction' source term is proportional
to the r.m.s. density fluctuations almost independent of scale.
Compared to the density fluctuations it amounts to a factor of about $70$
for SCDM.

\smallskip\noindent
$\bullet$ Using the ``Zel'dovich approximation'' we 
can calculate analytically the `backreaction' ${\cal B}_{\cal D}$.
This calculation shows that ${\cal B}_{\cal D}$ is a {\it growing} function
of time in an expanding universe.'' 

\bigskip

\section{BLUE's Conclusions}

``This debate brought up one fundamental result which supports RED: 
any inhomogeneous Newtonian
cosmology, whose flat space sections are confined to a length--scale $L$ on
which the matter variables are periodic, averages out to the standard model.
On this global scale there is no `backreaction' and the cosmological parameters
of the homogeneous--isotropic solutions of Friedmann's equations are 
well--defined also for the average cosmology 
on that scale. Setting up simulations of large--scale structure
in this way is correct, and a global comoving coordinate system can be 
introduced to scale the whole cosmology. This validates the common way of
constructing inhomogeneous models of the Universe.

On the other hand, GREEN's arguments
initiate two well--justified ways of saying that this architecture
is ``forced'' due to the settings of a) excluding curvature of the space 
sections, and b) requiring periodic boundary conditions:

\smallskip
One way is to analyze the effect of `backreaction' on scales smaller than the
periodicity scale, and base this analysis upon the standard model, however, by 
extending the spatial size of periodicity to very large scales. 
The results obtained in the framework of a well--tested approximation scheme
are three--fold: first, they show the 
importance of the influence of the inhomogeneities on average properties of
a chosen spatial domain, {\it although} the ``forcing conditions'' bring the
effect down to zero on the scale $L$.
Second, the `backreaction' value is numerically small on large
scales, but it is {\it always} larger than the r.m.s. density fluctuations; in other
words: taking the amplitudes of density fluctuations serious (e.g. by 
normalizing the cosmogony on some large scale) always implies the presence
of, e.g., shear fluctuations which are neglected on that scale 
in the standard model. 
Third, they show that the effect is
a growing function of time. From the latter result we may establish the 
notion of 
{\it global gravitational instability} of the standard model as opposed to
the well--known local instability: it states that, as soon as the `backreaction'
has a non--zero value at some time, this value will be increasing; the
average model drifts away from the standard model.

\smallskip
The second ``forcing'' is due to the Newtonian treatment. Being justified 
on smaller scales, a Newtonian model is expected to fail just when 
we approach the {\it large} scales of periodicity which we have to consider to 
justify neglection of the  `backreaction' effect. 
Setting up a general--relativistic model unavoidably
implies
the presence of local curvature and will, in general, yield a nonvanishing
average curvature for inhomogeneous models. Neither simple periodic boundary
conditions can be employed, nor can be proved that the `backreaction' effect
should vanish globally, at least for compact space sections without boundary.''

\section{Summary}

Both, RED and GREEN, are right, but if RED's assumptions are weakened,
the resulting cosmology has much richer properties and cannot be confined
to a simple box: it will take its additional degrees of freedom to evolve
away from the standard model.
GREEN's more general view suffers from the fact that an alternative model is 
yet not formulated, but it is definitely within reach.

\bigskip\bigskip

\acknowledgments

I would like to thank Martin Kerscher, Herbert Wagner 
and Arno Wei{\ss} for discussions, Arno Wei{\ss} for his
allowance to present Fig.1 prior to publication of a common
work. Some aspects of this contribution were subject of a letter
correspondence with J\"urgen Ehlers, to whom I am also thankful for 
discussing the manuscript.

This work was supported by the ``Sonderfor\-schungs\-bereich {\bf 375} f\"ur
Astro--Teilchenphysik der Deutschen Forschungsgemeinschaft''.

\end{document}